\documentclass[twocolumn,twoside,prl]{revtex4}

\usepackage{amsmath}
\usepackage{amssymb}
\usepackage{latexsym}
\usepackage{revsymb}
\usepackage{epsfig}

\newcommand{\ket}[1]{\left| #1 \right\rangle} 
\newcommand{\bra}[1]{\left\langle #1 \right|}
\newcommand{\av}[1]{\left\langle #1 \right\rangle}
\newcommand{\identity}{\openone}
\newcommand{\op}[1]{\hat{#1}}

\newcommand{\prior}{{\widetilde{P}}}

\begin{document}

\title{No quantum advantage for nonlocal computation}

\author{Noah Linden$^1$}
\author{Sandu Popescu$^{2,3}$}
\author{Anthony J. Short$^2$}
\author{Andreas Winter$^1$}

\affiliation{$^1$ Department of Mathematics, University of Bristol, %
 University Walk, Bristol BS8 1TW, U.K.}
\affiliation{$^2$ H.H.Wills Physics Laboratory, University of Bristol, %
 Tyndall Avenue, Bristol BS8 1TL, U.K.}
\affiliation{$^3$ Hewlett-Packard Laboratories, Stoke Gifford, %
 Bristol BS12 6QZ, U.K.}

\begin{abstract}
We investigate the problem of ``nonlocal" computation, in which separated parties must compute a function with nonlocally encoded inputs and output, such that each party individually learns nothing, yet together they compute the correct function output. We show that the best that can be done classically is a trivial linear approximation.  Surprisingly, we also show that quantum entanglement provides no advantage over the classical case. On the other hand, generalized (i.e. super-quantum) nonlocal correlations allow  perfect nonlocal computation. This gives new insights into the nature of quantum nonlocality and its relationship to generalised nonlocal correlations.
\end{abstract}

\maketitle

In 1964, John Bell~\cite{bell} proved that quantum theory can generate
correlations unachievable by any local classical means. These nonlocal correlations
cannot be used to transmit information, but are nevertheless useful for 
many information-theoretic tasks, including cryptography~\cite{E91},
dense-coding~\cite{BW92}, quantum teleportation~\cite{teleport},
and reducing communication complexity~\cite{comm-compl1, comm-compl2}.
Identifying those tasks which can benefit from the use of 
quantum nonlocality, and those which can not, is crucial to assessing and
understanding fully the power of quantum information processing.

Here, we report an unexpected limitation of quantum correlations
by exhibiting a large class of nonlocal tasks for which quantum
resources are of no benefit at all over local classical
strategies, even though the perfect execution of the task would 
not violate the non-signalling principle. These tasks can be
described in a unified way as the nonlocal computation of
Boolean functions.

Following \cite{fault-tolerance}, in which the problems of `nonlocal equality' and `nonlocal majority' are considered,  we define the nonlocal computation of a general Boolean function $f$ as follows.

Consider a Boolean function $f$ from $n$ bits $\{z_1,z_2,...,z_n\}$ to a single bit:
\begin{equation}
c=f(z_1,z_2,...,z_n).
\end{equation}
We now distribute each input bit to two parties, Alice and Bob, so that 
neither party individually learns anything about the global input \footnote{Note the distinction here between nonlocal computation and distributed computation. In the latter, some input bits are given to Alice and the rest to Bob, and hence each party learns something about the global input}. For each input bit, Alice is given a bit $x_i$, and Bob a bit $y_i$, such that their XOR is equal to $z_i$ ($z_i=x_i\oplus y_i$). However, individually $x_i$ and $y_i$ are totally random, being with equal probability 0 or 1.
To successfully perform the nonlocal computation, Alice must produce an output bit $a$ and Bob an output bit $b$ (without communicating with each other), such that $c=a\oplus b$. i.e. 
\begin{equation} 
a \oplus b =f(x_1 \oplus y_1, x_2 \oplus y_2,...,x_n \oplus y_n).
\end{equation}

The task we consider in this letter is for Alice and Bob to maximize the probability of success of their nonlocal computation, given some prior distribution on the inputs $z_i$ (for example 
the prior distribution could be for each $z_i$ to be 0 or 1 with equal probability) and either (a)  quantum resources, (b) classical resources alone, or (c) generalised non-signalling resources.  

Surprisingly, we find that quantum resources provide no advantage over classical resources for nonlocal computation. In fact both are very ineffective - the best they can do is just a trivial linear approximation of the computation. Since non-linearity is the essential element of computation, we could say that nonlocal computation is impossible in classical and quantum theory. 
This is particularly surprising because generalised non-signalling correlations \cite{PR, boxes} (including ``super-quantum'' correlations which  violate Bell inequalities by more than quantum theory) would allow perfect success in any nonlocal computation. This shows that nonlocal correlations in general \emph{are} helpful in such tasks, and that our results indicate a characterising feature of quantum nonlocality.

At the end of the Letter we generalize the situation
to more parties and to more general nonlocal tasks.

\medskip\noindent
{\bf Nonlocal computation with quantum resources.} Consider the nonlocal computation of a general Boolean function $f$ as above, given quantum resources. To simplify the notation, we denote the inputs by bit-strings $x=x_1x_2...x_n$, $y=y_1y_2...y_n$ and $z=z_1z_2...z_n$, and the bitwise-XOR by $\oplus$. The nonlocal computation of $c=f(z)$ can then be written as
\begin{equation} \label{eqn:f}
a \oplus b =  f(x \oplus y).
\end{equation}

Let us suppose that the inputs $z$ are given according to 
an arbitrary probability distribution $\prior(z)$  According to the above definition, 
to ensure that Alice or Bob alone have no individual
knowledge of $z$, it is necessary to take all inputs $x$ and $y$ satisfying $z=x\oplus y$ 
with equal probability. This means that each party individually
has a maximally random bit-string, with the joint probability distribution
for their inputs given by
\begin{equation}
  P(x,y) = \frac{1}{2^n} \prior(x \oplus y). 
\end{equation}
The average success probability for Alice and Bob to satisfy eq.~(\ref{eqn:f}) is therefore given 
in terms of the success probability for given inputs, $P(a\oplus b = f(x\oplus y)|xy)$, by

\begin{equation}
  P(f) = \frac{1}{2^n} \sum_{xy} \prior(x \oplus y) P(a\oplus b = f(x\oplus y)|xy) 
\end{equation}

In the most general quantum protocol, Alice and Bob share an entangled
quantum state $\ket{\psi}$ and perform projective measurements on their
subsystem dependant on their inputs, given by Hermitian operators
$\op{a}_x$ and $\op{b}_y$ respectively, with eigenvalues $0$ and $1$.
They then output their measurement results. Note that protocols involving
initially mixed states or POVM measurements can all be represented in this 
form by expanding the dimensionality of the initial state. In this quantum case,
\begin{equation}
  P(a\oplus b \!=\! f(x\oplus y)|xy)
            \!=\! \frac{1}{2} \left(\! 1 \!+\! \bra{\psi}\! (-1)^{f(x \oplus y)
                                              + \op{a}_x + \op{b}_y} \!\ket{\psi} \!\right) \!.
\end{equation}
Hence the total probability of success is given by 
\begin{equation} \label{eqn:PQf}
  P_Q(f) = \frac{1}{2} + \frac{1}{2^{n+1}}
                         \sum_{xy} \! \prior(x \oplus y) \bra{\psi}\!
                                      (-1)^{f(x \oplus y) + \op{a}_x + \op{b}_y} \!\ket{\psi}. 
\end{equation}
To further analyze $P_Q(f)$, we note that we can re-express it mathematically
in terms of a single scalar product in a larger Hilbert space (Note that
this does not correspond to any physical change, but is merely intended to 
 aid in the analysis). Extending the Hilbert space from $\mathcal{H}$ to
$\mathcal{H} \otimes \mathbb{C}^{2^n}$, we define normalised states $\ket{\alpha}$
and $\ket{\beta}$ and a Hermitian operator $\op{\Phi}$ as follows: 
\begin{align}
  \ket{\alpha} &= \frac{1}{\sqrt{2^{n}}} \sum_x (-1)^{\op{a}_x}\otimes \identity
                                         \ket{\psi} \otimes \ket{x},
  \label{eqn:alpha}                                                \\ 
  \ket{\beta}  &= \frac{1}{\sqrt{2^{n}}} \sum_y (-1)^{\op{b}_y}\otimes \identity
                                         \ket{\psi} \otimes \ket{y},
  \label{eqn:beta}                                                 \\ 
  \op{\Phi}    &= \sum_{xy} (-1)^{f(x \oplus y)} \prior(x \oplus y)
                                         \ket{x}\!\!\bra{y},
  \label{eqn:Phi}
\end{align}
where $\ket{x}$ and $\ket{y}$ are computational basis states in 
$\mathbb{C}^{2^n}$.
Eq.~(\ref{eqn:PQf}) can then be re-expressed in the simple form
\begin{equation} \label{eqn:qsimple} 
  P_Q(f) = \frac{1}{2}\bigl( 1 + \bra{\alpha} \identity \otimes \op{\Phi} \ket{\beta} \bigr),
\end{equation} 
from which it follows that
\begin{equation}\label{eqn:PQf2}
P_Q(f) \leq \frac{1}{2}\!
            \left( 1 \!+\! \bigl| \bra{\alpha} \bigr| \,
                            \bigl\|\identity \otimes \op{\Phi} \bigr\| \, 
                           \bigl| \ket{\beta} \bigr|             \right)
       =    \frac{1}{2}\! \left( 1 \!+\! \bigl\| \op{\Phi} \bigr\| \right) \!,
\end{equation}
where $\bigl\| \op{\Phi} \bigr\|$ is the operator norm of 
$\op{\Phi}$ (the largest modulus eigenvalue).

To investigate the eigenstates and eigenvalues of $\op{\Phi}$,
we first rewrite it in the Fourier-transform basis:
\begin{equation}
  \label{eq:tilde-u}
  \ket{\tilde{u}} = \frac{1}{\sqrt{2^{n}}}  \sum_x (-1)^{u .\, x} \ket{x} 
\end{equation}
where $u.x$ is the inner product modulo 2 of the bit strings
$u$ and $x$. This gives
\begin{equation}\begin{split}
  \op{\Phi} &= \left( \sum_u \ket{\tilde{u}}\!\!\bra{\tilde{u}} \right) 
                \op{\Phi} \left( \sum_v \ket{\tilde{v}}\!\!\bra{\tilde{v}} \right)              \\
            &= \frac{1}{2^n}\! \sum_{uvxy}\! (-1)^{f(x\oplus y) + u.x + v.y }
                                          \prior(x \oplus y) \ket{\tilde{u}}\!\!\bra{\tilde{v}} \\ 
            &= \frac{1}{2^n}\! \sum_{uvyz}\! \left(\! (-1)^{f(z) + u.z} \prior(z) \!\right)\!
                                         \left(\! (-1)^{(u+v).y} \!\right)
                                         \ket{\tilde{u}}\!\!\bra{\tilde{v}}                     \\ 
            &= \sum_u \left(\! \sum_z (-1)^{f(z) + u.z} \prior(z) \!\right)
                      \ket{\tilde{u}}\!\!\bra{\tilde{u}},
\end{split}\end{equation}
where in the second line we have replaced the sum over $x$ by one 
over $z=x \oplus y$. The eigenstates of $\op{\Phi}$ are therefore
$\ket{\tilde{u}}$, and inserting the modulus of the largest eigenvalue
in (\ref{eqn:PQf2}) we obtain the quantum bound
\begin{equation}
  P_Q(f) \leq  \frac{1}{2}\left( 1+\max_u \biggl| \sum_z (-1)^{f(z) + u.z} \prior(z) \biggr|
                           \right).
  \label{eqn:qbound}
\end{equation}

\medskip\noindent
{\bf Nonlocal computation with classical resources.}
We now consider the optimal classical strategy for performing the
distributed computation of $f$. Without loss of generality we restrict
our analysis to deterministic strategies, as the success probabilities
for random strategies will simply be a convex combination of these.

To analyse a general deterministic strategy, we simply replace the
operators $\op{a}_x$ and $\op{b}_y$ above with numbers $a_x$ and $b_y$ which represent
the outputs given by Alice and Bob for different inputs. The analogues of eqns.~(\ref{eqn:alpha}), (\ref{eqn:beta}) and  (\ref{eqn:qsimple}) 
for the classical success probability are then 
\begin{eqnarray}
  \ket{\alpha_c} &=& \frac{1}{\sqrt{2^{n}}} \sum_x (-1)^{a_x} \ket{x}, \\
  \ket{\beta_c}  &=& \frac{1}{\sqrt{2^{n}}} \sum_y (-1)^{b_y} \ket{y},
\\  \label{eqn:alphabetac}
P_C(f) &=& \frac{1}{2}\big( 1 +  \bra{\alpha_c}  \op{\Phi} \ket{\beta_c} \big)
\end{eqnarray}
By choosing the classical strategy
\begin{equation} \label{eqn:2party}
a_x = u.x \oplus \delta, \qquad b_y = u.y,
\end{equation}
where
\begin{equation} \label{eqn:2partydelta}
\delta = \begin{cases}
            1 & \text{ if } \sum_z (-1)^{f(z) + u.z} \prior(z) < 0, \\
            0 & \text{ otherwise,}
         \end{cases}
\end{equation}
we obtain
\begin{equation}\begin{split}
  P_C(f) &= \frac{1}{2} \left(\! 1 \!+\! \frac{1}{2^n}
                                         \sum_{xy} (-1)^{f(x \oplus y) + u.x+ u.y + \delta}
                                                   \prior(x \oplus y )                \!\right) \\
         &= \frac{1}{2} \left( 1 \!+\! \biggl| \sum_z (-1)^{f(z) + u.z} \prior(z) \biggr| \right),
\end{split}\end{equation}
and with the appropriate choice of $u$ we can therefore reach the
quantum bound given by (\ref{eqn:qbound}). Hence we have 
proved that quantum theory provides no advantage over the best classical 
strategy for the distributed computation of $f$.  Moreover, the solution 
described in (\ref{eqn:2party}) and (\ref{eqn:2partydelta}) is simply 
the best linear approximation of $f(z)$. 

\medskip\noindent
{\bf Nonlocal computation using generalized non-signalling correlations.}
In \cite{PR}, Popescu and Rohrlich asked whether or not quantum mechanics is uniquely determined by the existence of nonlocal correlations that are consistent with relativity (i.e. that do not allow signalling). Surprisingly they found that the class of possible non-signalling correlations is larger than the quantum mechanical one. The question then arises whether or not such correlations exist in nature and if not, why not. A great deal of research has been undertaken recently into such generalised non-signalling correlations \cite{boxes, van-dam, fault-tolerance} with the aim of better characterizing the differences between them and quantum correlations.
  
It is interesting to speculate whether it would be possible to beat the quantum bound for nonlocal computation using such generalised non-signalling correlations.  The answer is, quite trivially, yes.  Indeed, all that we require from the correlations is that they yield both possible sets of outputs fulfilling (\ref{eqn:f}) with equal probability (e.g. $a=0~ b=0 \; [50\%]\; a=1,~b=1\; [50 \%]$ when $f(x\oplus y)=0$). Each party individually will then obtain a random bit and learn nothing about the other party's input. Such a correlation is therefore non-signalling and fulfils (\ref{eqn:f}) perfectly, giving $\max P_G(f) = 1$.  (Here the index $G$ stands for ``generalized correlations".)

Incidentally, it is also easy to see that (except in the case when $f$ is constant)  for any generalized  correlation that fulfils (\ref{eqn:f}) perfectly, the local bits $a$ and $b$ have to be uniformly random to ensure non-signalling.

\medskip\noindent
{\bf Example: Nonlocal computation of AND.}
Probably the simplest non-trivial case is that of the AND function
\begin{equation} \label{eqn:AND}
  \text{AND}(z_1, z_2) = z_1 z_2. 
\end{equation}
The nonlocally distributed version of the AND function is given by \cite{fault-tolerance}
\begin{equation}
  a \oplus b = (x_1 \oplus y_1)(x_2 \oplus y_2),
  \label{eqn:dist-AND}
\end{equation}
where $x_1$ and $x_2$ are Alice's input bits, $y_1$ and $y_2$ are Bob's 
input bits, and $a$ and $b$ are Alice and Bob's respective output bits.

When the different values of the input bits $z_1$ and $z_2$ are given with uniform probability ($\prior(z)=\frac{1}{4}$), it is easy to show from (\ref{eqn:qbound}) that 
 \begin{equation} \nonumber
  P_C^{\max}(\text{AND}) =P_Q^{\max}(\text{AND}) =\frac{3}{4} < P_G^{\max}(\text{AND})=1.
\end{equation}

A simple classical strategy that achieves this bound is for  Alice and Bob to both give the output
zero in all cases ($u=\delta=0$). This strategy will only fail when
$x_1 \oplus y_1= x_2 \oplus y_2 =1$, which corresponds to $1/4$ of the
possible inputs, thus we obtain $\max P_C(\text{AND})=3/4$. 

Note the distinction between the nonlocal computation given by (\ref{eqn:dist-AND}) and the distributed computation represented by 
\begin{equation} \label{eqn:PR} 
a \oplus b = x_1  y_1.
\end{equation}
for which it can be shown that 
 \begin{equation} \nonumber 
  P_C^{\max} = \frac{3}{4} < P_Q^{\max} =\frac{2+\sqrt{2}}{4} <  P_G^{\max}=1.
\end{equation}
The classical and quantum bounds in this case correspond to the Clauser-Horne-Shimony-Holt \cite{chsh} and Tsirelson \cite{tsirelson} bounds respectively, and the generalised correlations satisfying (\ref{eqn:PR}) are commonly referred to as a PR-box \cite{PR, boxes}. Note that (\ref{eqn:PR}) does not correspond to the nonlocal computation of any $f(z)$.

It is possible to simulate nonlocal-AND perfectly using two PR-boxes and local operations \cite{fault-tolerance}. Interestingly, when the PR-boxes are made increasingly noisy, they yield a success probability of P(AND)=3/4, precisely when the noisy correlations would be attainable in quantum theory. 

\medskip\noindent
{\bf Discussion.}
The nonlocal version of some functions can be computed perfectly with a
local classical strategy (e.g. nonlocal-NOT can be implemented with
$u=\delta=1$); in fact, it is precisely the affine linear functions 
(modulo $2$) that can be implemented perfectly.
For all other cases, 
\begin{equation}
   P_C^{\max}(f) = P_Q^{\max}(f) < P_G^{\max}(f) = 1. 
\end{equation}
Quantum entanglement therefore offers no benefit over a local classical
strategy for the nonlocal computation of Boolean functions.
By contrast, generalised non-signalling correlations would allow perfect success 
in any such task. An interesting question is whether all super-quantum correlations are helpful in computing some distributed function. If this were indeed the case, we  would obtain a powerful and intuitive characterisation of quantum nonlocality.

Note that in the definition of success probability above
we assumed a fixed prior distribution $\prior$. One could equally well ask
for the maximum success probability \emph{in the worst case} (i.e. when each 
strategy is evaluated using its worst $z$). Fortunately the minimax theorem of 
game theory \cite{vonNeumann:Morgenstern} tells us that in the classical case,
when Alice and Bob can use shared randomness to access mixed strategies,  the
optimal worst-case success probability  is equal to the maximal success
probability $(P_C^{max}$) for some particular fixed prior distribution $\prior$. 
That quantum strategies can do no better then follows 
from the fact that $P_Q^{max} = P_C^{max}$ for the chosen $\prior$. Hence even 
in this scenario the identity of classical and quantum optimal performance is preserved.

It is straightforward to extend the results obtained above to distributed computations of $f(z)$ by any number of parties. 
In the general multi-party case, the function's inputs and output are encoded in the modulo 2 sum of 
$m$ separate inputs $x_i$ and outputs $a_i$ (with all sets of $x_i$ consistent with $z$ equally probable). 
Note that this task cannot be easier than distributed computation with only two parties, as the 
$m$-party case can be obtained from the 2-party case by encoding $y$ randomly in the modulo 2 sum of 
$m-1$ bits and then separating them. 
Hence the bounds on quantum and classical success probabilities obtained above must still apply to the $m$-party case. 
Furthermore, it is easy to see that the classical strategy 
\begin{equation} 
a_i(x_i) = \left\{ \begin{array}{ccl} u.x_i + \delta &:& i=0 \\ u.x_i &:& i=1 \ldots (m-1) \end{array}  \right.
\end{equation}
achieves the same success probability as the 2-party strategy given above, and therefore reaches the quantum bound given by (\ref{eqn:qbound}) for the optimal choice of $u$. 
 
Note that each choice of $f(z)$ and $\prior(z)$ corresponds to a Bell-type inequality: 
\begin{equation} \label{eq:gen-bell} 
\sum_{x,y} C(x,y) \av{A_x B_y}  \leq K
\end{equation}
where $A_x$ are $B_y$ are measurements with outcomes $\pm 1$, 
\begin{eqnarray} 
C(x,y) &=& (-1)^{f(x \oplus y)} \prior(x \oplus y), \\
K &=& 2^n (2 \max P_C(f) -1 ).
\end{eqnarray}  
Our results imply that there is also a Tsirelson-type inequality with exactly the same coefficients constraining the allowed quantum states. It would be interesting to discover if any of these inequalities generate facets of the Bell-polytope of classically attainable probability distributions $P(ab|xy)$ \cite{polytope}(and consequently a facet of the set of attainable quantum probability distributions). In any case, we find that the Bell-polytope and the (convex) Tsirelson-body have
many (potentially lower-dimensional) faces in common which are not
trivially inherited from the probability or non-signalling constraints. 

This analysis also leads us to a very considerable generalization of the nonlocal tasks
described so far.  Let us consider any Bell expression of the form
\begin{equation}
  \sum_{x,y} M(x,y) \av{A_x B_y}.
\end{equation} 
The matrix $M(x,y)$ need not be a function of $x \oplus y$ (as has been the case so far);
indeed it need not even be symmetric.  As long as the largest singular value of $M$
corresponds to an operator $\ket{\tilde{u}}\!\!\bra{\tilde{v}}$ 
with Hadamard basis vectors $\ket{\tilde{u}}$, $\ket{\tilde{v}}$
as in~(\ref{eq:tilde-u}), then quantum resources do not offer a benefit
over classical ones in performing the task.  While this gives a very wide class 
for which quantum mechanics provides no benefit, it is also worth pointing out that not
all nonlocal tasks for which this is true are of this type (see for example \cite{NL-SP-pq-games}).

Although functions with a single-bit output are very important (as they encapsulate all decision problems), it would also be interesting to extend these results to functions with a multi-bit output, or with different input and output alphabets (e.g. ternary rather than binary). In both cases, it is important to consider how success will be measured, as in addition to the total success probability considered above one could reasonably measure success by the average `distance' between the output and the correct answer. For functions with a multi-bit output, where success is measured by the number of correct output bits, our results immediately imply that quantum strategies provides no advantage over classical strategies (because the best strategy is to optimally compute each output bit independently).

\medskip\noindent
{\bf Acknowledgments.}
We thank Harry Buhrman for helpful conversations,
in particular for posing the question of the worst-case performance.
We also thank the U.K.~EPSRC for support through the ``QIP IRC'',
and the EC for support through the QAP project (contract
no.~IST-2005-15848). AW additionally acknowledges support
from a University of Bristol Research Fellowship.

\end{document}